\documentclass[aip,preprint]{revtex4-2}
\usepackage[cp1251]{inputenc}
\usepackage[T2A]{fontenc}
\usepackage[english]{babel}
\usepackage{amssymb,latexsym,amsmath,amscd}
\usepackage{graphicx,color,framed}
\usepackage{tikz}
\usepackage{xcolor}
\usepackage{siunitx}
\usepackage{empheq}
\usepackage{xr}
\usepackage{microtype}
\usepackage{amsfonts}
\usepackage[normalem]{ulem}
\graphicspath{{figures/}}

\makeatletter
\newcommand*{\addFileDependency}[1]{% argument=file name and extension
  \typeout{(#1)}
  \@addtofilelist{#1}
  \IfFileExists{#1}{}{\typeout{No file #1.}}
}
\makeatother

\begin{document}
\newcommand{\solid}{\raisebox{2pt}{\tikz\draw[solid](0,0) -- (5mm,0);}}

\title{Noether's second theorem and covariant field theory of mechanical stresses in inhomogeneous ionic liquids}
\author{\firstname{Petr E.} \surname{Brandyshev}}
%\homepage[]{Your web page}
%\thanks{}
%\altaffiliation{}
\affiliation{School of Applied Mathematics, HSE University, Tallinskaya st. 34, 123458 Moscow, Russia}

\author{\firstname{Yury A.} \surname{Budkov}}
\email[]{ybudkov@hse.ru}
%\homepage[]{Your web page}
%\thanks{}
%\altaffiliation{}
\affiliation{School of Applied Mathematics, HSE University, Tallinskaya st. 34, 123458 Moscow, Russia}
\affiliation{G.A. Krestov Institute of Solution Chemistry of the Russian Academy of Sciences, 153045, Akademicheskaya st. 1, Ivanovo, Russia}

\begin{abstract}
In this paper, we present a covariant approach that utilizes Noether's second theorem to derive a symmetric stress tensor from the grand thermodynamic potential functional. We focus on the practical case where the density of the grand thermodynamic potential is dependent on the first and second coordinate derivatives of the scalar order parameters. Our approach is applied to several models of inhomogeneous ionic liquids that consider electrostatic correlations of ions or short-range correlations related to packing effects. Specifically, we derive analytical expressions for the symmetric stress tensors of the Cahn-Hilliard-like model, Bazant-Storey-Kornyshev model, and Maggs-Podgornik-Blossey model. All of these expressions are found to be consistent with respective self-consistent field equations.
\end{abstract}

\maketitle
\section{Introduction}
Ionic liquids, e.g. electrolyte solutions and room temperature ionic liquids have gained significant attention from researchers and chemical engineers. This was mainly motivated by the use of ionic liquids in different applications including batteries, fuel cells, supercapacitors, lipid membranes, ion exchange resins, charged colloids, and many others. In all these examples, ionic liquids that either interact with charged surface of a membrane, macromolecule, colloid, or electrode surface, or are confined in the electrified nanopores are strongly inhomogeneous. Due to the liquid inhomogeneity, its description requires solution of the certain self-consistent field equations\cite{blossey2023poisson,budkov2021electric} for electrostatic potential and local ionic concentrations with appropriate boundary conditions.\par
In practice, for ionic liquids in nano-confinement of an arbitrary geometry (in nanosized pores, for instance), it is important to calculate the mechanical stresses described by the {\sl stress tensor}~\cite{shi2023perspective}, in addition to concentration and electrostatic potential profiles. Knowledge of the local stress tensor, which is consistent with the self-consistent field equations, would allow us to calculate important physical quantities such as solvation pressure and shear stresses~\cite{kolesnikov2021models,kolesnikov2022electrosorption,gurina2022disjoining}. These quantities, in turn, provide us with the ability to estimate the deformation of the pore material, characterized by a certain elastic modulus~\cite{kolesnikov2021models}, which is relevant for batteries and supercapacitors, where the microporous electrodes impregnated by the liquid electrolytes are extensively used (see, for instance,~\cite{li2018high,da2020reviewing,chen2022porous,koczwara2017situ}). On the other hand, the stress tensor allows one to calculate the macroscopic force acting on the charged macroscopic conductor or dielectric immersed in the ionic liquid~\cite{budkov2022modified,budkov2023macroscopic}.\par Thereby, a first-principle approach that enables us to derive the stress tensor of the inhomogeneous ionic liquids would be valuable for applications.\par

Budkov and Kolesnikov made the first attempts in this direction in ref.\cite{budkov2022modified}. They applied Noether's theorem to the grand thermodynamic potential of ionic liquid as a functional of the electrostatic potential and established the conservation law, $\partial_{i}\sigma_{ik}=0$, which means the local mechanical equilibrium condition expressed via the symmetric stress tensor, $\sigma_{ik}$. The obtained stress tensor consists of two terms: Maxwell electrostatic stress tensor, related to the local electric field, and isotropic hydrostatic stress tensor, which is determined by the local osmotic pressure of the ions. It is important to note that the authors generalized the local mechanical equilibrium condition in presence of the external potential forces acting on the ions. Based on it, they derived the general analytical expression for the electrostatic disjoining pressure of ionic liquid confined in a slit charged nanopore, generalizing the well-known DLVO expression for the case of an arbitrary reference model of liquid. In ref.\cite{budkov2023macroscopic} Budkov and Kalikin presented a self-consistent field theory of macroscopic forces in inhomogeneous flexible chain polyelectrolyte solutions. Subjecting the system to the small dilation and taking into account the self-consistent field equations following from the extremal condition of the grand thermodynamic potential, the authors derived an analytical expression for a stress tensor. The latter, in addition to the aforementioned hydrostatic and Maxwell stress tensors, contains a stress tensor, arising from the conformational entropy of flexible polymer chains, -- conformational stress tensor. The authors applied their theory to the description of polyelectrolyte solutions confined in a conducting slit nanopore and observed anomalous behavior of disjoining pressure and electric differential capacitance at sufficiently small pore thickness. \par

However, in the aforementioned examples, the authors only considered local mean-field theories, without taking into account the electrostatic correlations between ions and their short-range correlations related to packing effects. These effects are encompassed by a nonlocal free energy functional, which in its simplest form may be reduced to the functionals containing second derivatives of the order parameters (i.e. ionic concentrations and electrostatic potential)~\cite{bazant2011double,misra2019theory,de2020continuum,blossey2017structural,ciach2018simple,de2020interfacial}. Thus, it is essential to refine the approach based on Noether's theorem for deriving the stress tensors from the free energy functionals. Moreover, it is highly advantageous for this methodology to be able to examine the curved geometries, described by arbitrary curvilinear coordinate systems through a general covariant formulation utilizing Noether's theorem. To the best of our knowledge, no such approach has been proposed in current literature.\par

Noether's theorems~\cite{noether1971invariant} play a crucial role in different fields of modern physics, such as classical and quantum mechanics, classical and quantum field theory, and condensed matter physics~\cite{hermann2022noether,kourkoulou2022improved}. It is well known that Noether's first theorem enables us to derive conservation laws from the condition of invariance of certain functionals, such as action in classical field theory~\cite{bogolyubov1973introduction,landau1975classical,kourkoulou2022improved}, with respect to global transformations. Noether's second theorem is related to the concept of local symmetries, which holds a prominent position in the realms of gauge field theory, quantum field theory, and particle physics~\cite{bogolyubov1973introduction,peskin2018introduction}. Noether's second theorem establishes a connection between the symmetries of an action functional, in the field of mathematics and theoretical physics, and a differential equation system. The theorem specifies that if the action has an infinite-dimensional Lie algebra of infinitesimal symmetries parameterized linearly by $k$ arbitrary functions and their derivatives up to order $m$, the functional derivatives of action functional fulfill a system of $k$ differential equations~\cite{noether1971invariant}. Moreover, Noether's second theorem allows us to construct a functional which is invariant under the certain local transformations of fields~\cite{bogolyubov1973introduction,landau1975classical,kourkoulou2022improved}.\par

In this paper, we present a general covariant approach that is based on the application of Noether's second theorem. This approach allows us to derive a symmetric stress tensor from the grand thermodynamic potential functional, which contains first and second coordinate derivatives of the scalar order parameters. We apply this approach to various nonlocal models of ionic liquids, thereby deriving the expressions of their total stress tensors.

\section{Noether's second theorem for grand thermodynamic potential}
The grand thermodynamic potential (GTP) of a spatially inhomogeneous liquid expressed in arbitrary curvilinear coordinate system, depending on the scalar functions $\psi_{\alpha}(x)$ (order parameters)\footnote{In this work we consider only the case of scalar order parameters. The practically important case of a vector order parameter will be published elsewhere.} and their first and second derivatives and also on metric tensor elements $g_{ij}(x)$ and their first derivatives ($i,j=1,2,...,n$ -- coordinate indices in $n$-dimensional space, $dx$ -- volume element, $a$ -- number of a scalar function) can be written as follows\cite{landau1975classical,weinberg1972gravitation}
\begin{equation}\label{Omega}
\Omega[\psi,g]=\int_{V} dx \sqrt{g}\omega(\psi,\partial\psi,\partial\partial \psi,g_{ij},\partial g_{ij}),
\end{equation}
where $g=\det{g_{ij}}$. The form of the scalar functions $\psi_{\alpha}(x)$ for which the thermodynamic potential reaches an extreme value can be found from the self-consistent field equations, i.e. the Euler-Lagrange (EL) equations 
\begin{equation}\label{int1}
\bar{\delta}_{\psi}\Omega[\psi,g_{ij}]=0,
\end{equation}
Delta with a dash denotes the variation of potential only due to the variation of the form of functions $\psi$. It can be shown that 
\begin{equation}\label{3}
\bar{\delta}_{\psi}\tilde{\omega}(x)=\frac{\partial \tilde{\omega}}{\partial\psi_{\alpha}}\bar{\delta}\psi_{\alpha}+\frac{\partial \tilde{\omega}}{\partial\psi_{\alpha,i}}\bar{\delta}\psi_{\alpha,i}
+\frac{\partial \tilde{\omega}}{\partial\psi_{\alpha,ij}}\bar{\delta}\psi_{\alpha,ij},
\end{equation}
where the following designations have been introduced
\begin{equation}\label{}
\psi_{\alpha,i}=\partial_{i}\psi_{\alpha}(x),\quad \psi_{\alpha,ij}=\partial_{i}\partial_{j}\psi_{\alpha}(x),\quad \tilde{\omega}=\sqrt{g}\omega.
\end{equation}
Then, integrating (\ref{3}) by parts, we get
\begin{equation}\label{}
\begin{aligned}
\bar{\delta}_{\psi} \Omega=\int dx\bigg[\bigg(\frac{\partial\tilde{\omega}}{\partial\psi_{\alpha}}-\partial_{i}\bigg(\frac{\partial\tilde{\omega}}{\partial\psi_{\alpha,i}}\bigg)
+\partial_{i}\partial_{j}\bigg(\frac{\partial\tilde{\omega}}{\partial\psi_{\alpha,ij}}\bigg)\bigg)\bar{\delta}\psi_{\alpha}\\
+\partial_{i}\bigg(\frac{\partial\tilde{\omega}}{\partial\psi_{\alpha,i}}\bar{\delta}\psi_{\alpha}
-\partial_{j}\bigg(\frac{\partial\tilde{\omega}}{\partial\psi_{\alpha,ij}}\bigg)\bar{\delta}\psi_{\alpha}
+\frac{\partial\tilde{\omega}}{\partial\psi_{\alpha,ij}}\bar{\delta}\psi_{\alpha,j}\bigg)\bigg].
\end{aligned}
\end{equation}
In accordance with the well-known Einstein rule, repeated indices imply summation. If one requires that on the boundary, $G$, of the integration region, $V$, the variations of the functions $\psi$ and their derivatives are equal to zero, then from the extreme condition, i.e. the EL equations
\begin{equation}\label{stat}
\frac{\delta\Omega}{\delta\psi_{\alpha}}=0,
\end{equation}
where the variational (functional) derivative in this case takes the form
\begin{equation}\label{div1}
\frac{\delta\Omega}{\delta\psi_{\alpha}}=\frac{\partial\tilde{\omega}}{\partial\psi_{\alpha}}-\partial_{i}\bigg(\frac{\partial\tilde{\omega}}{\partial\psi_{\alpha,i}}\bigg)
+\partial_{i}\partial_{j}\bigg(\frac{\partial\tilde{\omega}}{\partial\psi_{\alpha,ij}}\bigg).
\end{equation}
The form of the functions $g_{ij}(x)$ depends on used curvilinear coordinate system (or on the manifold geometry~\footnote{This situation could arise, for instance, in the presence of an adsorption layer on a curved surface.} in general case). It is easy to check that if the functional form of $g_{ij}(x)$ changes, then the variation of the thermodynamic potential density takes the form
\begin{equation}\label{}
\bar{\delta}_{g}\tilde{\omega}(x)=\frac{\partial \tilde{\omega}}{\partial g_{ij}}\bar{\delta}  g_{ij}
+\frac{\partial \tilde{\omega}}{\partial g_{ij,k}}\bar{\delta}  g_{ij,k},
\end{equation}
Then, integrating by parts, one reduces the variation of the functional to the form 
\begin{equation}\label{5}
\bar{\delta}_{g} \Omega=\int dx\bigg[\frac{1}{2}\sqrt{g}T^{ij}\bar{\delta}g_{ij}
+\partial_{k}\bigg(\frac{\partial \tilde{\omega}}{\partial g_{ij,k}}\bar{\delta}  g_{ij}\bigg)\bigg]
\end{equation}
where the symmetric tensor $T^{ij}$ is defined by the equality
\begin{equation}\label{T}
\frac{1}{2}\sqrt{g}T^{ij}=\frac{\delta \Omega}{\delta g_{ij}},
\end{equation}
where variational derivative is
\begin{equation}\label{varder}
    \frac{\delta \Omega}{\delta g_{ij}}=\frac{\partial \tilde{\omega}}{\partial g_{ij}}
-\partial_{k}\bigg(\frac{\partial \tilde{\omega}}{\partial g_{ij,k}}\bigg).
\end{equation}
Now let us consider the following diffeomorphic transformations
\begin{equation}\label{diff1}
x^{k'}=x^{k}+\xi^{k}(x),
\end{equation}
\begin{equation}\label{diff2}
\delta g^{ik}(x)=\xi^{i,k}+\xi^{k,i},
\end{equation}
\begin{equation}\label{}
\bar{\delta} g^{ik}(x)=\xi^{i;k}+\xi^{k;i},
\end{equation}
\begin{equation}\label{g}
\bar{\delta} g_{ik}(x)=-\xi_{i;k}-\xi_{k;i},
\end{equation}
where
\begin{equation}\label{}
\delta g^{ik}(x)=g^{i'k'}(x')-g^{ik}(x),
\end{equation}
\begin{equation}\label{}
\bar{\delta} g^{ik}(x)=g^{i'k'}(x)-g^{ik}(x)
\end{equation}
and $\xi^{i;k}$ is the covariant derivative of the vector $\xi^{i}$ (see, for instance, Refs. \cite{weinberg1972gravitation,landau1975classical}), i.e.
\begin{equation}
\label{}
\xi^{i}_{~;k}=\nabla_{k}\xi^{i}=\partial_{k}\xi^{i}+\Gamma^{i}{}_{jk}\xi^{j},
\end{equation}
where $\Gamma^{i}{}_{jk}=g^{im}\left(g_{mk,j}+g_{mj,k}-g_{kj,m}\right)/2$ is the standard Christoffel symbols~\cite{weinberg1972gravitation,landau1975classical}.\par
\bigskip

Let the functional (\ref{Omega}) be invariant under diffeomorphisms (\ref{diff1} -- \ref{diff2}). As is well known, if some functional, introduced in Euclidean space, is invariant under global shifts and rotations, then one can ensure it to be an invariant under local transformations (diffeomorphisms), formally substituting covariant derivatives with the partial derivatives and using the metric tensor to raise the indices~\cite{landau1975classical}. In what follows, we will omit the index $\alpha$ over which the summation in equation (\ref{3}) is implied. \par
Let us consider variations of functions under above diffeomorphisms
\begin{equation}\label{}
\delta\psi(x)=\psi'(x')-\psi(x),
\end{equation}
\begin{equation}\label{varpsi}
\bar{\delta}\psi(x)=\psi'(x)-\psi(x)=\delta\psi(x)-\xi^{k}\partial_{k}\psi(x),
\end{equation}
Due to the scalar nature of these functions, we have
\begin{equation}\label{4}
\delta\psi(x)=0,~\bar{\delta}\psi(x)=-\xi^{k}\partial_{k}\psi(x).
\end{equation}
Variation of the functional in general form is
\begin{equation}\label{}
\delta \Omega=\delta\int dx \tilde{\omega}(x)=\int dx' \tilde{\omega}'(x')-\int dx \tilde{\omega}(x).
\end{equation}
The variation of the integrand in expression
\begin{equation}\label{}
\tilde{\omega}'(x')=\tilde{\omega}(\psi'(x'),g_{i'j'}(x'))=\tilde{\omega}(x)+\delta\tilde{\omega}(x),
\end{equation}
can be divided into two parts
\begin{equation}\label{}
\delta\tilde{\omega}(x)=\delta_{\psi}\tilde{\omega}(x)+\delta_{g}\tilde{\omega}(x),
\end{equation}
where
\begin{equation}\label{}
\delta_{g}\tilde{\omega}(x)=\frac{\partial \tilde{\omega}}{\partial g_{ij}}\delta  g_{ij}
+\frac{\partial \tilde{\omega}}{\partial g_{ij,k}}\delta  g_{ij,k},
\end{equation}
\begin{equation}\label{}
\delta_{\psi}\tilde{\omega}(x)=\frac{\partial \tilde{\omega}}{\partial\psi}\delta\psi+\frac{\partial \tilde{\omega}}{\partial\psi_{,i}}\delta\psi_{,i}
+\frac{\partial \tilde{\omega}}{\partial\psi_{,ij}}\delta\psi_{,ij},
\end{equation}
On the other hand, we can write
\begin{equation}\label{}
\delta\tilde{\omega}(x)=\bar{\delta}\tilde{\omega}(x)+\xi^{i}\partial_{i}\tilde{\omega},
\end{equation}
where $\bar{\delta}\tilde{\omega}(x)$ is the variation of $\tilde{\omega}(x)$ due to a change in a functional form similar to (\ref{varpsi}). Then we can write
\begin{equation}\label{}
\bar{\delta}\tilde{\omega}(x)=\bar{\delta}_{\psi}\tilde{\omega}(x)+\bar{\delta}_{g}\tilde{\omega}(x).
\end{equation}
The functional variation of the first order in $\xi^{k}$ is
\begin{equation}\label{1}
\delta \Omega=\int dx\left(\bar{\delta}\tilde{\omega}(x)+\xi^{i}\partial_{i}\tilde{\omega}\right)+\int dx'\tilde{\omega}(x)-\int dx\tilde{\omega}(x).
\end{equation}
Then, using the expression
\begin{equation}\label{}
dx'=\left(1+\partial_{k} \xi^{k}\right)dx,
\end{equation}
we can write identity
\begin{equation}\label{2}
\int dx'\tilde{\omega}(x)-\int dx\tilde{\omega}(x)=\int dx \partial_{k} \xi^{k}\tilde{\omega}(x)
\end{equation}
from which substituting (\ref{2}) into (\ref{1}), we can get
\begin{equation}\label{}
\delta \Omega=\int dx\left[\bar{\delta}\tilde{\omega}(x)+\partial_{i}\bigg(\tilde{\omega}(x)\xi^{i}(x)\bigg)\right],
\end{equation}
Introducing the notation
\begin{equation}\label{}
\delta_{\psi} \Omega=\int dx\left[\bar{\delta}_{\psi}\tilde{\omega}(x)+\partial_{i}\bigg(\tilde{\omega}(x)\xi^{i}(x)\bigg)\right],
\end{equation}
the total variation of the functional can be written in the form
\begin{equation}\label{DeltaOmega}
\delta\Omega=\delta_{\psi}\Omega+\bar{\delta}_{g}\Omega.
\end{equation}
Integrating in eq. (\ref{3}) by parts again, we arrive at
\begin{equation}\label{}
\begin{aligned}
\delta_{\psi} \Omega=\int dx\bigg[\bar{\delta}\psi\frac{\delta\Omega}{\delta\psi}+\partial_{i}\bigg(\frac{\partial\tilde{\omega}}{\partial\psi_{,i}}\bar{\delta}\psi-\partial_{j}\bigg(\frac{\partial\tilde{\omega}}{\partial\psi_{,ij}}\bigg)\bar{\delta}\psi\\
+\frac{\partial\tilde{\omega}}{\partial\psi_{,ij}}\bar{\delta}\psi_{,j}+\tilde{\omega}(x)\xi^{i}(x)\bigg)\bigg],
\end{aligned}
\end{equation}
Then, using the equality (\ref{4}), we can obtain
\begin{equation}\label{7}
\begin{aligned}
{\delta}_{\psi} \Omega=-\int dx\bigg[\xi^{k}\psi_{,k}\frac{\delta\Omega}{\delta\psi}
+\partial_{i}\bigg(\frac{\partial\tilde{\omega}}{\partial\psi_{,i}}\psi_{,k}\xi^{k}-\partial_{j}\bigg(\frac{\partial\tilde{\omega}}{\partial\psi_{,ij}}\bigg)\psi_{,k}\xi^{k}\\
+\frac{\partial\tilde{\omega}}{\partial\psi_{,ij}}\partial_{j}(\psi_{,k}\xi^{k})-\tilde{\omega}(x)\xi^{i}\bigg)\bigg],
\end{aligned}
\end{equation}
Similarly, rewriting (\ref{5}) with account for (\ref{g}), we obtain
\begin{equation}\label{}
\bar{\delta}_{g} \Omega=-\int dx\bigg[\sqrt{g}T^{ij}\xi_{i;j}
+2\partial_{k}\bigg(\frac{\partial \tilde{\omega}}{\partial g_{ij,k}}\xi_{i;j}\bigg)\bigg]
\end{equation}
Integrating by parts, we get
\begin{equation}\label{Omega_g}
\bar{\delta}_{g} \Omega=\int dx\bigg[\sqrt{g}\nabla_{j}T^{ji}\xi_{i}
-\partial_{k}\bigg(\sqrt{g}T^{kj}\xi_{j}+2\frac{\partial \tilde{\omega}}{\partial g_{ij,k}}\xi_{i;j}\bigg)\bigg]
\end{equation}
The invariance of the GTP functional requires that its variation is equal to zero regardless of the functional form of $\xi_{i}(x)$, so that we can always choose them and their first derivatives to be equal to zero on the boundary of integration region, $G$.  Then, using the divergence theorem, we can show that the integrals from the total divergences contained in the right-hand sides of (\ref{7}) and (\ref{Omega_g}) are equal to zero. Therefore, summing up (\ref{7}) and (\ref{Omega_g}), we can show that from the invariance condition of the functional under diffeomorphisms (\ref{diff1}--\ref{diff2})
\begin{equation}\label{}
\delta\Omega={\delta}_{\psi}\Omega+\bar{\delta}_{g}\Omega=0
\end{equation}
and due to the independence and arbitrariness of the functions $\xi_{i}(x)$ we can get equation
\begin{equation}\label{theorem}
\sqrt{g}\nabla_{j}T^{jk}=\psi^{,k}\frac{\delta\Omega}{\delta\psi},
\end{equation}
which is nothing but a special case of the Noether's second theorem. The condition of stationarity,
$\delta\Omega/\delta\psi=0$, of the functional (\ref{Omega}) leads to the $"$conservation law$"$ 
\begin{equation}\label{theorem_2}
\nabla_{j}T^{jk}=0, 
\end{equation}
which can be interpreted as a local mechanical equilibrium condition of the liquid with symmetric stress tensor $T_{ik}$, expressed in an arbitrary curvilinear coordinate system. In Euclidean space, where $g_{ij}=\delta_{ij}$ we have
\begin{equation}\label{T2}
\sigma^{ij}=T^{ij}\bigg|_{g_{ij}=\delta_{ij}}=2\frac{\delta\Omega }{\delta g_{ij}}\bigg|_{g_{ij}=\delta_{ij}}.
\end{equation}
At the small deformations, $\xi_{i}$, the metric tensor is $g_{ij}=\delta_{ij}+2u_{ij}$, where $u_{ij}=(\xi_{i,j}+\xi_{j,i})/2$ is the strain tensor~\cite{landau1986theory}, one can get the expression obtained for the first time for polyelectrolyte solutions in~\cite{budkov2023macroscopic}
\begin{equation}
\sigma^{ij}=\frac{\delta\Omega }{\delta u_{ij}}\bigg|_{u_{ij}=0}.
\end{equation}

In the next section, subjecting the GTP to the most general diffeomorphisms, we will establish an explicit expression of the stress tensor $\sigma_{ij}$.

\section{General local transformations}
Now let us consider the most general diffeomorphic transformations when neither $\xi_{k}$ nor their derivatives are equal to zero at the integration boundary. The method outlined below will enable us to derive an alternative expression for the stress tensor $T^{ik}$ in terms of the GTP density derivatives with respect to $\psi$, $\psi_{,i}$, and $\psi_{,ij}$. In this case, the variation (\ref{DeltaOmega}) can be rewritten as follows
\begin{equation}\label{}
\delta\Omega=-\int dx \bigg[B^{k}\xi_{k}(x)+B^{ji}\xi_{i,j}(x)+B^{ijk}\xi_{i,jk}(x)\bigg],
\end{equation}
then, due to the arbitrariness of $\xi_{k}(x)$, this variation is zero if the following equalities
\begin{equation}\label{A1}
B^{ijk}=\frac{\partial \tilde{\omega}}{\partial \psi_{,kj}} \psi^{,i}+\frac{\partial \tilde{\omega}}{\partial g_{ij,k}}+\frac{\partial \tilde{\omega}}{\partial g_{ik,j}}=0,
\end{equation}
\begin{equation}\label{A2}
\begin{aligned}
B^{ji}=-\sqrt{g}J^{ji}-\frac{\partial \tilde{\omega}}{\partial \psi_{,jm}}\Gamma^{i}{}_{mn}\psi^{,n}+\partial_{n}\bigg(\frac{\partial \tilde{\omega}}{\partial \psi_{,nj}} \psi^{,i}\bigg)\\
+2\frac{\partial \tilde{\omega}}{\partial g_{,ij}}
-2\frac{\partial \tilde{\omega}}{\partial g_{mn,j}}\Gamma^{i}{}_{mn}=0,
\end{aligned}
\end{equation}
\begin{equation}\label{A3}
\begin{aligned}
B^{i}=\psi^{,i}\frac{\delta\Omega}{\delta\psi}-\partial_{j}\bigg(\sqrt{g}J^{ji}\bigg)
-\partial_{j}\bigg(\frac{\partial \tilde{\omega}}{\partial \psi_{,jm}}\Gamma^{i}{}_{mn}\psi^{,n}\bigg)\\
-2\frac{\partial \tilde{\omega}}{\partial g_{,mn}} \Gamma^{i}{}_{mn}
-2\frac{\partial \tilde{\omega}}{\partial g_{mn,j}} \partial_{j}\Gamma^{i}{}_{mn}=0
\end{aligned}
\end{equation}
are satisfied, where we have introduced the notation for the tensor $J^{ik}$ which is defined by expression
\begin{equation}\label{J}
\sqrt{g}J^{ik}=-\frac{\partial\tilde{\omega}}{\partial\psi_{,i}}\psi^{,k}
+\partial_{j}\bigg(\frac{\partial\tilde{\omega}}{\partial\psi_{,ij}}\bigg)\psi^{,k}\\
-\frac{\partial\tilde{\omega}}{\partial\psi_{,ij}}\nabla_{j}(\psi^{,k})
+\tilde{\omega}g^{ik}
\end{equation}
Despite the presence of a partial derivative in the right-hand side of eq. (\ref{J}), $J^{ik}$ is indeed a tensor (see Appendix A). \par
It can be shown that equation (\ref{A1}) turns into an identity using Christoffel symbols definition (for more details, see Appendix $B$). In fact, the first $n^3$ equations define the well-known dependence of the Christoffel symbols $\Gamma^{i}{}_{jk}$ on the metric tensor $g_{ij}$ and its first coordinate derivatives, if we initially assume $\Gamma^{i}{}_{jk}$ to be unknown.\par

Using eq. (\ref{A1}), we can express the derivatives of the thermodynamic potential density with respect to $g_{ij,k}$ via linear combinations of the derivatives with respect to $\psi_{,ij}$ (see equality (\ref{B4}) in Appendix $B$). Performing these calculations and substituting the result into eq. (\ref{A2}), we can reduce eq. (\ref{A2}) to the form (see Appendix C)
\begin{equation}\label{TJ}
T^{ij}=J^{ij}+\nabla_{k} \bigg(A^{ikj}-A^{kij}\bigg),
\end{equation}
where 
\begin{equation}\label{Ag}
\sqrt{g}A^{ikj}=\frac{\partial \tilde{\omega}}{\partial \psi_{,jk}} \psi^{,i}.
\end{equation}
Note that due to the fact that $T^{ij}$ is symmetric, the right hand side of eq. (\ref{TJ}) is also symmetric with respect to $i$ and $j$.\par
It is easy to check, due to the antisymmetry of tensor $A^{ikj}-A^{kij}$ with respect to $i$ and $k$, that the following equality
\begin{equation}\label{}
\nabla_{j}T^{ji}=\nabla_{j}J^{ji}
\end{equation}
is satisfied. Eq. (\ref{TJ}) allows us to consider $T^{ij}$ as the stress tensor, because, as a consequence of Noether's (first) theorem~\cite{bogolyubov1973introduction}, the $"$current$"$ $J^{ij}$ satisfies the conservation law due to the invariance of the GTP with respect to the global shift transformations (see below).
\par
If we take the divergence, $\partial_{j}$, from eq. (\ref{A2}) and then subtract eq. (\ref{A3}), with the use of eq. (\ref{A1}) on the solutions of the EL equations, we arrive at eq. (\ref{theorem}).\par

For Euclidean space we can obtain
\begin{equation}\label{9}
j^{ik}=J^{ik}\bigg|_{g_{ij}=\delta_{ij}}=-\frac{\partial \omega}{\partial\psi_{,i}}\psi^{,k}+
\partial_{j}\bigg(\frac{\partial \omega}{\partial\psi_{,ij}}\bigg)\psi^{,k}
+\omega\delta^{ik}-\frac{\partial \omega}{\partial\psi_{,ij}}\psi^{,jk}.
\end{equation}
Thus, using eq. (\ref{theorem}), we obtain
\begin{equation}\label{8}
\partial_{i}j^{ik}=\psi^{,k}\frac{\delta\Omega}{\delta\psi}
\end{equation}
We can see that tensor $j_{ik}$, in contrast to (\ref{T}), is not symmetric. However, it satisfies the conservation law, $\partial_{i}j^{ik}=0$, on the solutions of the EL equations (\ref{stat}). In particular case of the GTP invariance with respect to the global shift transformations in Euclidean space, $j^{ik}$ can be interpreted as the stress tensor up to the divergence of the third rank tensor which is antisymmetric with respect to the couple of indices, i.e.
\begin{equation}\label{}
\sigma^{ik}=j^{ik}+\partial_{l}(a^{ilk}-a^{lik}),
\end{equation}
where the second term is divergenceless due to the antisymmetry in the indices $i$ and $j$. Using eq. (\ref{Ag}) at $g_{ij}=\delta_{ij}$, we arrive at 
\begin{equation}\label{}
a^{ilk}=\frac{\partial \omega}{\partial\psi_{,lk}}\psi^{,i}.
\end{equation}
Then tensor $\sigma^{ik}$ takes the form
\begin{equation}\label{sigma_2}
\begin{aligned}
\sigma^{ik}=\omega\delta^{ik}-\frac{\partial \omega}{\partial\psi_{,i}}\psi^{,k}+
\partial_{j}\bigg(\frac{\partial \omega}{\partial\psi_{,ij}}\bigg)\psi^{,k}\\
-\partial_{j}\left(\frac{\partial \omega}{\partial\psi_{,ik}}\psi^{,j}\right)
+\partial_{j}\bigg(\frac{\partial \omega}{\partial\psi_{,jk}}\bigg)\psi^{,i}
+\frac{\partial \omega}{\partial\psi_{,kj}}\psi^{,ji}-\frac{\partial \omega}{\partial\psi_{,ij}}\psi^{,jk}.
\end{aligned}
\end{equation}
Using eq. (\ref{TJ}) and taking into account that 
\begin{equation}\label{identity}
\frac{1}{2}\bigg(\frac{\partial \omega}{\partial\psi_{,i}}\psi^{,k}-\frac{\partial \omega}{\partial\psi_{,k}}\psi^{,i}\bigg)+
\frac{\partial \omega}{\partial\psi_{,ij}}\psi^{,jk}-\frac{\partial \omega}{\partial\psi_{,kj}}\psi^{,ji}=0,
\end{equation}
we arrive at the final expression for the stress tensor in explicitly symmetric form
\begin{equation}\label{sigma}
\begin{aligned}
\sigma^{ik}=-\frac{1}{2}\bigg(\frac{\partial \omega}{\partial\psi_{,i}}\psi^{,k}+\frac{\partial \omega}{\partial\psi_{,k}}\psi^{,i}\bigg)
+\omega\delta^{ik}
-\partial_{j}\left(\frac{\partial \omega}{\partial\psi_{,ik}}\psi^{,j}\right)\\
+\partial_{j}\bigg(\frac{\partial \omega}{\partial\psi_{,ij}}\bigg)\psi^{,k}
+\partial_{j}\bigg(\frac{\partial \omega}{\partial\psi_{,kj}}\bigg)\psi^{,i}.
\end{aligned}
\end{equation}
Note that eq. (\ref{identity}) is simply follows from eq. (\ref{TJ}). Eq. (\ref{sigma}) is the main result of this paper.

For the case when the integrand in (\ref{Omega}) does not depend on the second coordinate derivatives of the order parameters, as it follows from (\ref{sigma_2}), stress tensor simplifies to
\begin{equation}
\label{sigma_local}
\sigma^{ik}=\omega\delta^{ik}-\psi^{,i}\frac{\partial \omega}{\partial\psi_{,k}}.
\end{equation}
Expression (\ref{sigma_local}) obtained for the first time in ref.~\cite{budkov2022modified} within different approach.

Note that in one dimensional case the mechanical equilibrium condition, $\partial_{i}\sigma_{ik}=0$, simplifies to 
\begin{equation}
\frac{d\sigma_{xx}}{dx}=0
\end{equation}
or
\begin{equation}
\sigma_{xx}=const.
\end{equation}
The latter expression is simply a first integral of the self-consistent field equations~\cite{blossey2022comprehensive,blossey2017structural,budkov2016theory,budkov2021theory,budkov2020two}.

\section{Stress tensors for ionic liquid models}
In this section we will apply the formulated above formalism to establish the analytical expressions for the stress tensors which are realized within the different self-consistent field theories of the spatially inhomogeneous ionic liquids. Note that this formalism can be also applied to simple non-ionic liquids. We will not consider here the pure mean-field theory (or the modified Poisson-Boltzmann theory) considered by one of us recently in ref.\cite{budkov2022modified}. We will focus only on the theories taking into account molecular structure or electrostatic correlations of the ions beyond pure mean-field theory.
\subsection{Cahn-Hilliard-like model}
Firstly, we consider the ionic liquid model with account for the so-called structural interactions within the Cahn-Hilliard~\cite{cahn1965phase} approach, within which the molecular structure effects are taken into account via the quadratic form over the concentration gradients~\cite{cahn1965phase,blossey2017structural}. Thus, the GTP of the ionic liquid in this model can be written in the form 
\begin{equation}
\Omega = F_{el}+\Omega_{liq},
\end{equation}
where
\begin{equation}
F_{el}=\int d^{3}x\left(-\frac{\varepsilon \left(\nabla\psi\right)^2}{2}+\rho\psi\right)
\end{equation}
is the electrostatic free energy of the ionic liquid with the electrostatic potential $\psi(x)$; the local charge density of ions is $\rho(x)=\sum_{\alpha}q_{\alpha}c_{\alpha}(x)$, $q_{\alpha}$ is the electrostatic charge of ion of the $\alpha^{th}$ type; $\varepsilon$ is the dielectric permittivity of medium. The GTP of the reference liquid system is
\begin{equation}
\label{omega_liquid}
\Omega_{liq} = \int d^{3}x\left(f(\{c_{\alpha}\})+\frac{1}{2}\sum\limits_{\alpha\gamma}\kappa_{\alpha\gamma}\nabla c_{\alpha}\cdot\nabla c_{\gamma} -\sum\limits_{\alpha}\mu_{\alpha}c_{\alpha}\right),
\end{equation}
where the first term in the integrand is the free energy density of liquid as the function of the local ionic concentrations, $c_{\alpha}$, in the local density approximation; the second term is the contribution of the so-called structural interactions of the ions within the Cahn-Hilliard approach with the structural constants $\kappa_{\alpha\gamma}$ which are proportional to the bulk correlation lengths\cite{cahn1965phase,blossey2017structural}; $\mu_{\alpha}$ are the bulk chemical potentials of the species. Thus, the GTP can be rewritten in the form
\begin{equation}
\label{omega_liquid_2}
\Omega =\int d^{3}x\omega(x),
\end{equation}
where the GTP density is
\begin{equation}
\omega=-\frac{\varepsilon \left(\nabla\psi\right)^2}{2}+\rho\psi+f(\{c_{\alpha}\})+\frac{1}{2}\sum\limits_{\alpha\gamma}\kappa_{\alpha\gamma}\nabla c_{\alpha}\cdot\nabla c_{\gamma}-\sum\limits_{\alpha}\mu_{\alpha}c_{\alpha}.
\end{equation}
 Note that integration in (\ref{omega_liquid_2}) is performed over the volume of the ionic liquid bounded by the surfaces of conducting or dielectric macroscopic bodies. The EL equations, 
\begin{equation}
\frac{\partial \omega}{\partial{\psi}}-\partial_{i}\left(\frac{\partial\omega}{\partial\psi_{,i}}\right)=0,~\frac{\partial \omega}{\partial{c_{\alpha}}}-\partial_{i}\left(\frac{\partial\omega}{\partial c_{\alpha,i}}\right)=0,
\end{equation}
yield
\begin{equation}
\label{EL_CH}
\nabla^2\psi = -\frac{1}{\varepsilon}\sum\limits_{\alpha}q_{\alpha}c_{\alpha},~\sum\limits_{\gamma}\kappa_{\alpha\gamma}\nabla^2 c_{\gamma} = v_{\alpha}, 
\end{equation}
where $v_{\alpha}=q_{\alpha}\psi+\partial{f}/\partial{c_{\alpha}}-\mu_{\alpha}$. The first equation is the standard Poisson equation for the electrostatic potential, whereas the second one describes the thermodynamic equilibrium condition for ions of the $\alpha^{th}$ kind. The system of these self-consistent field equations can be solved with the appropriate boundary conditions for the electrostatic potential and local ionic concentrations\footnote{The boundary conditions for the Poisson equation are determined by the nature of the macroscopic bodies immersed in the ionic liquid. For the local ionic concentrations we can assume $c_{\alpha}=0$ at the surface of the immersed bodies. The latter boundary condition  reflects the fact that in the vicinity of the surface each ion undergoes a strong repulsive force.}. Thereby, using eq. (\ref{sigma_local}) and excluding the bulk chemical potentials from the final result using eq. (\ref{EL_CH}), we arrive at
\begin{equation}
\sigma_{ik}=\sigma_{ik}^{(M)}+\sigma_{ik}^{(h)}+\sigma_{ik}^{(s)},
\end{equation}
where 
\begin{equation}
\label{Maxwell}
\sigma_{ik}^{(M)}=\varepsilon \bigg(\partial_{i}\psi\partial_{k}\psi-\frac{1}{2}\delta_{ik}(\nabla\psi)^{2}\bigg) 
\end{equation}
is the Maxwell stress tensor,
\begin{equation}
\label{hyd}
\sigma_{ik}^{(h)}=-P\delta_{ik}    
\end{equation}
is the hydrostatic stress tensor with the local osmotic pressure, $P=\sum_{\alpha}c_{\alpha}\partial{f}/\partial{c_{\alpha}}-f$, of the reference liquid system, whereas
\begin{equation}
\sigma_{ik}^{(s)}=\sum\limits_{\alpha\gamma}\sigma^{(\alpha\gamma)}_{ik}
\end{equation}
is the {\sl structural} contribution to the total stress tensor with the following auxiliary tensors
\begin{equation}
\sigma_{ik}^{(\alpha\gamma)}=\kappa_{\alpha\gamma}\left[\left(\frac{1}{2}\nabla c_{\alpha}\cdot \nabla c_{\gamma} +c_{\alpha}\nabla^2c_{\gamma}\right)\delta_{ik}-\partial_{i}c_{\alpha}\partial_{k}c_{\gamma}\right].
\end{equation}

\subsection{Bazant-Storey-Kornyshev model}
Let us move to the discussion  of the Bazant-Storey-Kornyshev (BSK) model for which the GTP density can be written in the form
\begin{equation}
\omega=-\frac{\varepsilon}{2}\left(\left(\nabla\psi\right)^2+l_{c}^2\left(\nabla^2\psi\right)^2\right)+\rho\psi+f(\{c_{\alpha}\})-\sum\limits_{\alpha}\mu_{\alpha}c_{\alpha},
\end{equation}
where the first term, in contrast to the classical modified Poisson-Boltzmann theory~\cite{budkov2022modified}, contains additional "nonlocal term" $-\varepsilon l_c^2(\nabla^2\psi)^2/2$, related to the electrostatic correlations of ions beyond the mean-field approximation, where $l_c$ is the correlation length\cite{bazant2011double,de2020continuum,misra2019theory}. Note that $l_{c}$ can be also interpreted as the quadrupolar length of the medium~\cite{slavchov2014quadrupole,slavchov2014quadrupole_2,budkov2020statistical,avni2020charge}. In original BSK theory~\cite{bazant2011double} the free energy density is described by the one that is realized for the symmetric lattice gas model, although in practical calculations one can use other reference liquid models, such as hard sphere mixture or ideal gas mixture~\cite{budkov2022modified,de2020continuum}. The EL equations have the following form
\begin{equation}
\label{EL}
\nabla^2\psi-l_c^2\Delta^2\psi = -\frac{1}{\varepsilon}\sum\limits_{\alpha}q_{\alpha}c_{\alpha},~\frac{\partial{f}}{\partial{c_{\alpha}}}=\mu_{\alpha}-q_{\alpha}\psi.
\end{equation}
Using eq. (\ref{sigma}), we arrive at the following stress tensor
\begin{equation}
\sigma_{ik}=\sigma_{ik}^{(h)}+\sigma_{ik}^{(el)},
\end{equation}
where $\sigma_{ik}^{(h)}$ is determined by eq. (\ref{hyd}) and electrostatic stress tensor consists of two contributions
\begin{equation}
\sigma_{ik}^{(el)}=\sigma_{ik}^{(M)}+\sigma_{ik}^{(cor)},
\end{equation}
where $\sigma_{ik}^{(M)}$ is determined by eq. (\ref{Maxwell}) and the electrostatic correlation contribution to the total stress tensor is
\begin{equation}
\label{sigmacor}
\sigma_{ik}^{(cor)}=\varepsilon l_{c}^2\left(\left(\bold{E}\cdot\nabla^2\bold{E}+\frac{1}{2}\left(\nabla\cdot\bold{E}\right)^2\right)\delta_{ik}-E_{i}\nabla^2 E_{k}-E_{k}\nabla^2 E_{i}\right),
\end{equation}
where $E_{i}=-\partial_{i}\psi$ are the electric field components. Eq. (\ref{sigmacor}) was recently obtained in paper~\cite{de2020continuum} within different approach.

\subsection{Maggs-Podgornik-Blossey model}
Now, let us consider the Maggs-Podgornik-Blossey (MPB) model~\cite{blossey2017structural} of the spatially inhomogeneous ionic liquid. The GTP of ionic liquid in the Euclidean space within the MPB model has the form
\begin{equation}
\Omega = \int d^{3}x\left(-\frac{\varepsilon \left(\nabla\psi\right)^2}{2}+\rho\psi+f\left(\{\bar{c}_{\alpha}\}\right)-\sum\limits_{\alpha}\mu_{\alpha}c_{\alpha}\right),
\end{equation}
where the first and second terms in the integrand, as above, describe the electrostatic energy density of ionic liquid and the third term is the intrinsic free energy density of ionic liquid in the weighted density approximation which is dependent on the weighted concentrations  
\begin{equation}
\bar{c}_{\alpha}(x)=\hat{w}_{\alpha}c_{\alpha}(x)=\int d^{3}x^{\prime}{w}_{\alpha}(|x-x^{\prime}|)c_{\alpha}(x^{\prime})
\end{equation}
with the self-adjoint integral operator, $\hat{w}_{\alpha}$, with the kernel (weighting function) $w_{\alpha}(|x-x^{\prime}|)$.

Now, let us reorganize the GTP expression in the following manner
\begin{equation}
\Omega = -\int d^{3}x\frac{\varepsilon \left(\nabla\psi\right)^2}{2}+\int d^{3}x\left(f\left(\{\bar{c}_{\alpha}\}\right)-\sum\limits_{\alpha}\bar{\mu}_{\alpha}\hat{w}_{\alpha}^{-1}\bar{c}_{\alpha}\right),
\end{equation}
where 
\begin{equation}
\bar{\mu}_{\alpha}=\mu_{\alpha}-q_{\alpha}\psi.
\end{equation}
Taking into account that due to self-adjointness of operator $\hat{G}_{\alpha}=\hat{w}_{\alpha}^{-1}$ the following equality~\cite{landau2013quantum}
\begin{equation}
\int d^{3}x\bar{\mu}_{\alpha}\hat{G}_{\alpha}\bar{c}_{\alpha}=\int d^{3}x \bar{c}_{\alpha}\hat{G}_{\alpha}\bar{\mu}_{\alpha},
\end{equation}
is hold and using the Legendre transform\cite{maggs2016general,podgornik2018general,podgornik2018general,budkov2018theory,budkov2016theory,budkov2021theory}
\begin{equation}
P(\{\mu_{\alpha}\})=\sum\limits_{\alpha}c_{\alpha}\mu_{\alpha}-f(\{c_{\alpha}\}),~\mu_{\alpha}=\frac{\partial f}{\partial c_{\alpha}},
\end{equation}
we get
\begin{equation}\label{omega}
\Omega[\psi]=-\int d^{3}x  \bigg(\frac{\varepsilon \left(\nabla\psi\right)^2}{2}+P(\{\mu_{\alpha}-q_{\alpha}\bar{\psi}_{\alpha}\})\bigg),
\end{equation}
where we have determined the following weighted potentials
\begin{equation}
\bar{\psi}_{\alpha}=\hat{G}_{\alpha}\psi.
\end{equation}
and $P(\{\mu_{\alpha}\})$ is the local pressure of liquid which is the function of the chemical potentials, $\mu_{\alpha}$, of species.
Using the following model weighting functions
\begin{equation}
w_{\alpha}(|x-x^{\prime}|)=\frac{1}{4\pi l_{\alpha}^2}\frac{e^{-\frac{|x-x^{\prime}|}{l_{\alpha}}}}{|x-x^{\prime}|},
\end{equation}
with some phenomenological lengths, $l_{\alpha}$, we obtain
\begin{equation}
\hat{G}_{\alpha}=1-l_{\alpha}^2\Delta.
\end{equation}
The phenomenological lengths, $l_{\alpha}$, determine the regions within which the local concentrations are weighed around certain point. To obtain the stress tensor, let us use eq. (\ref{T2}). For this purpose, the thermodynamic potential (\ref{omega}) can be rewritten in the following general covariant form
\begin{equation}\label{Omega2}
\Omega[\psi]=-\int d^{3}x \sqrt{g} \bigg(\frac{\varepsilon g^{ij}\partial_{i}\psi\partial_{j}\psi}{2}+P(\{\mu_{\alpha}-q_{\alpha}\bar{\psi}_{\alpha}\})\bigg),
\end{equation}
where Laplacian is~\cite{landau1975classical}
\begin{equation}
\label{b}
\Delta\psi=\frac{1}{\sqrt{g}}\partial_{i}\left(\sqrt{g}g^{ij}\partial_{j}\psi\right).
\end{equation}
Using the relations~\cite{landau1975classical}
\begin{equation}\label{}
\delta(\sqrt{g})=\frac{1}{2}\sqrt{g}g^{ij}\delta g_{ij},\quad \delta g^{ij}=-g^{im}g^{jn}\delta g_{mn},
\end{equation}
and varying the functional (\ref{Omega2}) with respect to metric tensor, we obtain
\begin{equation}\label{}
\delta\Omega=\frac{1}{2}\int d^{3}x \sqrt{g} \delta g_{ij}\bigg[\varepsilon \bigg(\partial^{i}\psi\partial^{j}\psi-\frac{1}{2}g^{ij}\partial_{n}\psi\partial^{n}\psi\bigg)-g^{ij}P\bigg]+\sum_{\alpha}\int d^{3}x \sqrt{g} \bar{c}_{\alpha} q_{\alpha} \delta\bar{\psi}_{\alpha},
\end{equation}
where 
\begin{equation}\label{}
\bar{c}_{\alpha}=\frac{\partial P}{\partial \mu_{\alpha}}
\end{equation}
are the weighted ionic concentrations, determined via the EL equation for the electrostatic potential.

Considering that $\partial^{i}=g^{ij}\partial_{j}$ and using eq. (\ref{b}), the variation of the GTP takes the form
\begin{equation}
\nonumber
\delta\Omega=\frac{1}{2}\int d^{3}x \sqrt{g} \delta g_{ij}\bigg[\varepsilon \bigg(\partial^{i}\psi\partial^{j}\psi-\frac{1}{2}g^{ij}\partial_{n}\psi\partial^{n}\psi\bigg)-g^{ij}P\bigg]
\end{equation}
\begin{equation}
\label{var2}
+\frac{1}{2}\sum_{\alpha}\bigg[l_{\alpha}^2q_{\alpha}\int d^{3}x \sqrt{g} \bar{c}_{\alpha} \Delta\psi g^{ij}\delta g_{ij}\bigg]-\sum_{\alpha}\bigg[\int d^{3}x
\bar{c}_{\alpha} \partial_{i}\bigg(\delta\bigg(\sqrt{g}g^{ij}\bigg)\partial_{j}\psi\bigg)\bigg],
\end{equation}
Integrating the last term by parts, using the divergence theorem and taking into account that at the boundary $\bar{c}_{\alpha}=0$, we obtain
\begin{equation}\label{}
\begin{aligned}
\delta\Omega = \frac{1}{2}\int d^{3}x \sqrt{g}\delta g_{ij}\bigg[\varepsilon \bigg(\partial^{i}\psi\partial^{j}\psi-\frac{1}{2}g^{ij}\partial_{n}\psi\partial^{n}\psi\bigg)-g^{ij}P\bigg]\\
+\frac{1}{2}\int d^{3}x \sqrt{g}\delta g_{ij}\sum_{\alpha} l_{\alpha}^2q_{\alpha} \bigg[g^{ij}\bigg(\bar{c}_{\alpha} \Delta\psi
+\partial_{n}\psi \partial^{n}\bar{c}_{\alpha}\bigg)-\partial^{i}\psi \partial^{j}\bar{c}_{\alpha}-\partial^{j}\psi \partial^{i}\bar{c}_{\alpha}\bigg],
\end{aligned}
\end{equation}
Stress tensor can be obtained from the expression
\begin{equation}\label{}
\delta\Omega=\frac{1}{2}\int d^{3}x \sqrt{g}T^{ij}\delta g_{ij}.
\end{equation}
Putting $g_{ij}=\delta_{ij}$, i.e. passing to the Euclidean space, in accordance with eq. (\ref{T2}) we arrive at
\begin{equation}
\label{sigma_tot}
\sigma_{ik}=\sigma_{ik}^{(M)}+\sigma_{ik}^{(h)}+\sigma_{ik}^{(cor)},
\end{equation}
where the first and second terms have already been determined above and
\begin{equation}
\label{stress1}
\begin{aligned}
\sigma_{ik}^{(cor)}=\sum_{\alpha} l_{\alpha}^2q_{\alpha} \bigg[\bigg(\bar{c}_{\alpha} \Delta\psi
+\nabla\psi \cdot \nabla \bar{c}_{\alpha}\bigg)\delta_{ik}-\partial_{i}\psi \partial_{k}\bar{c}_{\alpha}-\partial_{k}\psi \partial_{i}\bar{c}_{\alpha}\bigg]
\end{aligned}
\end{equation}
describes the contribution of the ionic correlations to the total stress tensor within the nonlocal density approximation. Note that the use of eq. (\ref{sigma}) brings us to the same analytical result (\ref{sigma_tot}).

\section{Conclusions}
In conclusion, we derived the expression for the stress tensor from the grand thermodynamic potential of an inhomogeneous liquid, using  a general covariant approach based on the application of Noether's second theorem. Our approach considers practically important cases where the grand thermodynamic potential density depends on the first and second coordinate derivatives of the order parameters. We have established two methods to obtain the symmetric stress tensor of the liquid: the first involves the functional differentiation of the grand thermodynamic potential functional in general covariant form with respect to the metric tensor (see eq. (\ref{T2})), while the second utilizes the explicit analytical expression (\ref{sigma}) derived using general diffeomorphic transformations. We have applied our approach to three models of the inhomogeneous ionic liquids: the Cahn-Hilliard-like model, the Bazant-Storey-Kornyshev model, and the Maggs-Podgornik-Blossey model. As a result, we have derived analytical expressions for the symmetric stress tensors that are consistent with corresponding self-consistent field equations.

%\textbf{Data availability statement.} {\sl The data that supports the findings of this study are available within the article and its
%supplementary material.}

\section*{Acknowledgments}
YAB thanks the Russian Science Foundation (Grant No. 22-13-00257) for financial support.

\section{Appendix A}
Let us introduce new variables
\begin{equation}\label{}
\varphi_{i}=\psi_{,i},
\end{equation}
\begin{equation}\label{t1}
\varphi_{ij}(\varphi_{n})=\psi_{,ij}-\Gamma^{m}{}_{ij}\varphi_{m}.
\end{equation}
Then the dependence on the variable $\varphi_{i}$ is defined by a complex function
\begin{equation}\label{}
\tilde{\omega}(\varphi_{i})=\tilde{\omega}(\varphi_{i},\varphi_{ij}(\varphi_{i})).
\end{equation}
Let us consider the expression
\begin{equation}\label{t}
\sqrt{g}T^{i}=\frac{\partial\tilde{\omega}}{\partial\psi_{,i}}
-\partial_{j}\bigg(\frac{\partial\tilde{\omega}}{\partial\psi_{,ij}}\bigg).
\end{equation}
Functional density depends on the covariant derivative with respect to $\psi_{,i}$, thus, the first term on the right-hand side of (\ref{t}) is not a tensor density because covariant derivative contains the non covariant part depending on $\psi_{,i}$ that is equal to the  contraction of $\psi_{,i}$ and Christoffel symbols (the second term on the right-hand side (\ref{t1})). After the introduction of new variables the derivative with respect to $\psi_{,i}$ can be obtained as the derivative of the composite function
\begin{equation}
\label{}
\frac{\partial\tilde{\omega}}{\partial\psi_{,i}}=\frac{d\tilde{\omega}}{d\varphi_{i}},
\end{equation}
where
\begin{equation}\label{}
\frac{d\tilde{\omega}}{d\varphi_{i}}=\frac{\partial\tilde{\omega}}{\partial\varphi_{i}}
+\frac{\partial\tilde{\omega}}{\partial\varphi_{mn}}\frac{\partial\varphi_{mn}}{\partial\varphi_{i}},
\end{equation}
Thus, substituting here (\ref{t1}), we obtain
\begin{equation}\label{t2}
\frac{\partial\tilde{\omega}}{\partial\psi_{,i}}=\frac{\partial\tilde{\omega}}{\partial\varphi_{i}}
-\frac{\partial\tilde{\omega}}{\partial\psi_{,mn}}\Gamma^{i}{}_{mn},
\end{equation}
It is obvious that the first term in the right hand side of (\ref{t2}) is the first rank tensor density.\par The second term in (\ref{t}) is not a tensor density, but eq. (\ref{t}) can be rewritten as follows
\begin{equation}\label{t3}
\sqrt{g}T^{i}=\frac{\partial\tilde{\omega}}{\partial\psi_{,i}}
-\sqrt{g}\nabla_{j}\bigg(\frac{\partial \omega}{\partial\psi_{,ij}}\bigg)+\frac{\partial\tilde{\omega}}{\partial\psi_{,mn}}\Gamma^{i}{}_{mn},
\end{equation}
where the second term is the tensor density, since the functional does not contain the covariant derivatives $\psi_{,ij}$ and, thereby, in the second term, under the covariant derivative, there is a second rank tensor. Substituting (\ref{t2}) into (\ref{t3}) and dividing by $\sqrt{g}$, we obtain the first rank tensor
\begin{equation}
\label{}
T^{i}=\frac{\partial \omega}{\partial\varphi_{i}}-\nabla_{j}\bigg(\frac{\partial \omega}{\partial\psi_{,ij}}\bigg).
\end{equation}
Indeed, both terms in the right hand side of this equation are tensors, so that $T^{i}$ is also tensor like (\ref{T}).

\section{Appendix B}
Let $g_{ij,k}$ to be included in $\tilde{\omega}$ only as a part of the covariant derivative, i.e.
\begin{equation}\label{B3}
\tilde{\omega}=\tilde{\omega}(\psi,\psi_{,i},\nabla_{s}\psi_{,l}).
\end{equation}
Covariant derivative has the form
\begin{equation}\label{}
\nabla_{s}\psi_{,l}=\psi_{,sl}-\tilde{\Gamma}^{m}{}_{sl}\psi_{,m}.
\end{equation}
Let us introduce the following notation
\begin{equation}\label{}
\phi_{sl}=\nabla_{s}\psi_{,l}.
\end{equation}
Then, we have
\begin{equation}\label{}
\frac{\partial \tilde{\omega}}{\partial \psi_{,mn}}=\frac{\partial \tilde{\omega}}{\partial \phi_{sl}} \frac{\partial \phi_{sl}}{\partial \psi_{,mn}},
\end{equation}
where
\begin{equation}\label{}
\frac{\partial \phi_{sl}}{\partial \psi_{,mn}}=\delta^{(m}{}_{s} \delta^{n)}{}_{l},
\end{equation}
i.e. we can write
\begin{equation}\label{B2}
\frac{\partial \tilde{\omega}}{\partial \psi_{,mn}}=\frac{\partial \tilde{\omega}}{\partial \phi_{sl}} \delta^{(m}{}_{s} \delta^{n)}{}_{l},
\end{equation}
while the derivative $g_{ij,k}$ in this case is
\begin{equation}\label{B1}
\frac{\partial \tilde{\omega}}{\partial g_{ij,k}}=-\frac{\partial \tilde{\omega}}{\partial \phi_{sl}} \frac{\partial \tilde{\Gamma}^{m}{}_{sl}}{\partial g_{ij,k}}\psi_{,m}.
\end{equation}
Using the definition of the Cristoffel symbols
\begin{equation}\label{Cristoffel}
\Gamma^{m}{}_{sl}=\frac{1}{2}g^{mn}(g_{ln,s}+g_{sn,l}-g_{sl,n}),
\end{equation}
we can obtain
\begin{equation}\label{}
\frac{\partial \Gamma^{m}{}_{sl}}{\partial g_{ij,k}}=g^{mn}(\delta^{(k}{}_{s} \delta^{j)}{}_{l} \delta^{i}{}_{n}+
\delta^{(k}{}_{s} \delta^{i)}{}_{l} \delta^{j}{}_{n}-\delta^{(i}{}_{s} \delta^{j)}{}_{l} \delta^{k}{}_{n}).
\end{equation}
Then, if we put
\begin{equation}\label{}
\tilde{\Gamma}^{m}{}_{sl}=\Gamma^{m}{}_{sl},
\end{equation}
eq. (\ref{B1}) takes the form
\begin{equation}\label{B5}
\frac{\partial \tilde{\omega}}{\partial g_{ij,k}}=-\frac{1}{2}\frac{\partial \tilde{\omega}}{\partial \phi_{sl}}
\bigg(\delta^{(k}{}_{s} \delta^{j)}{}_{l} \psi^{,i}+
\delta^{(k}{}_{s} \delta^{i)}{}_{l} \psi^{,j}-\delta^{(i}{}_{s} \delta^{j)}{}_{l} \psi^{,k}\bigg).
\end{equation}
Using eq. (\ref{B2}), the latter equation can be reduced to
\begin{equation}\label{B4}
\frac{\partial \tilde{\omega}}{\partial g_{ij,k}}=-\frac{1}{2}\left(\frac{\partial \tilde{\omega}}{\partial \psi_{,kj}} \psi^{,i}+
\frac{\partial \tilde{\omega}}{\partial \psi_{,ki}} \psi^{,j}-\frac{\partial \tilde{\omega}}{\partial \psi_{,ij}} \psi^{,k}\right)
\end{equation}
where taking into account symmetry $g_{ij}$ and $\psi_{,ij}$ in indices $i$ and $j$, we can obtain (\ref{A1}). To do this, it is necessary to swap $j$ and $k$ in the equation (\ref{B4}), and then add the resulting equation to the original one.\par 
Thus, we prove that from the definition of the Cristoffel symbols (\ref{Cristoffel}) we can obtain eq. (\ref{A1}). Obviously, we can proove that if the density of the functional can be written as (\ref{B3}), then, as it follows from (\ref{B4}), affine connection takes the form
\begin{equation}\label{}
\tilde{\Gamma}^{m}{}_{sl}=\frac{1}{2}g^{mn}(g_{ln,s}+g_{sn,l}-g_{sl,n})+S^{m}{}_{sl},
\end{equation}
where $S^{m}{}_{sl}$ is some arbitrary tensor which does not depend on $g_{ij,k}$. Since we initially assumed that the functional (\ref{Omega}) does not depend on any additional tensors, then we set $S^{m}{}_{sl}=0$.
\section{Appendix C}
Taking into account the symmetry of $\Gamma^{i}{}_{mn}$ in indices $m$ and $n$, one can get from eq. (\ref{B4})
\begin{equation}\label{}
-2\frac{\partial \tilde{\omega}}{\partial g_{mn,j}}\Gamma^{i}{}_{mn}=2\frac{\partial \tilde{\omega}}{\partial \psi_{,jn}} \psi^{,m}\Gamma^{i}{}_{mn}
-\frac{\partial \tilde{\omega}}{\partial \psi_{,mn}} \psi^{,j}\Gamma^{i}{}_{mn}.
\end{equation}
Substituting this identity to eq. (\ref{A2}), we get
\begin{equation}\label{C2}
\begin{aligned}
-\sqrt{g}J^{ji}+\frac{\partial \tilde{\omega}}{\partial \psi_{,jm}}\Gamma^{i}{}_{mn}\psi^{,n}+\partial_{n}\bigg(\frac{\partial \tilde{\omega}}{\partial \psi_{,nj}} \psi^{,i}\bigg)\\
+2\frac{\partial \tilde{\omega}}{\partial g_{,ij}}
-\frac{\partial \tilde{\omega}}{\partial \psi_{,mn}} \psi^{,j}\Gamma^{i}{}_{mn}=0.
\end{aligned}
\end{equation}
Using (\ref{B4}) again, we obtain the following expression
\begin{equation}\label{C1}
2\partial_{n}\bigg(\frac{\partial \tilde{\omega}}{\partial g_{ij,n}}\bigg)=
\partial_{n}\bigg(\frac{\partial \tilde{\omega}}{\partial \psi_{,ij}} \psi^{,n}-
\frac{\partial \tilde{\omega}}{\partial \psi_{,ni}} \psi^{,j}\bigg)
-\partial_{n}\bigg(\frac{\partial \tilde{\omega}}{\partial \psi_{,nj}} \psi^{,i}\bigg).
\end{equation}
Then, using (\ref{C1}) and (\ref{T}-\ref{varder}) we can obtain
\begin{equation}\label{}
2\frac{\partial \tilde{\omega}}{\partial g_{,ij}}=\sqrt{g}T^{ji}+\partial_{n}\bigg(\frac{\partial \tilde{\omega}}{\partial \psi_{,ij}} \psi^{,n}-
\frac{\partial \tilde{\omega}}{\partial \psi_{,ni}} \psi^{,j}\bigg)
-\partial_{n}\bigg(\frac{\partial \tilde{\omega}}{\partial \psi_{,nj}} \psi^{,i}\bigg).
\end{equation}
Thus, substituting the latter expression into eq. (\ref{C2}), we get
\begin{equation}
\label{C3}
\begin{aligned}
-\sqrt{g}J^{ji}+\sqrt{g}T^{ji}+\partial_{n}\bigg(\frac{\partial \tilde{\omega}}{\partial \psi_{,ij}} \psi^{,n}-
\frac{\partial \tilde{\omega}}{\partial \psi_{,ni}} \psi^{,j}\bigg)\\
+\Gamma^{i}{}_{mn}\frac{\partial \tilde{\omega}}{\partial \psi_{,jm}}\psi^{,n}
-\Gamma^{i}{}_{mn}\frac{\partial \tilde{\omega}}{\partial \psi_{,mn}} \psi^{,j}=0.
\end{aligned}
\end{equation}
We can reduce (\ref{C3}) to (\ref{TJ}) using the following identity
\begin{equation}\label{}
\sqrt{g}\nabla_{n}T^{nij}=\partial_{n}\bigg(\sqrt{g}T^{nij}\bigg)+\sqrt{g}\Gamma^{i}{}_{mn}T^{mnj}+\sqrt{g}\Gamma^{j}{}_{mn}T^{min},
\end{equation}
if we put 
\begin{equation}
\sqrt{g}T^{nij}=\sqrt{g}\bigg(A^{jni}-A^{nji}\bigg)=
\frac{\partial \tilde{\omega}}{\partial \psi_{,ni}} \psi^{,j}
-\frac{\partial \tilde{\omega}}{\partial \psi_{,ij}} \psi^{,n}.
\end{equation}

\bibliographystyle{aipnum4-2}
\bibliography{sample}

\end{document}